# Question Type, Cognitive Load, and CEFR Alignment: Evaluating LLM-Generated EFL Grammar Drill Exercises

**Steve WOOLLASTON[a*], Brendan FLANAGAN[b], Yuko TOYOKAWA[a], & Hiroaki OGATA[a]**
[a]*Kyoto University, Japan*
[b]*Ritsumeikan University, Japan*
*s.m.woollaston@gmail.com

**Abstract:** This study evaluates the pedagogical viability of LLM-generated English as a Foreign Language (EFL) learning content. Utilising log data from Japanese junior high school students practicing on a grammar drilling application, we analysed how different question modalities impact student performance and whether theoretical localised CEFR difficulty tiers accurately predict empirical task difficulty. Results reveal a clear performance hierarchy: multiple-choice questions carried the lowest cognitive load, cloze tasks posed the greatest barrier to active recall, and drag-and-drop exercises incurred the heaviest time penalties. Furthermore, learner data validated the CEFR-J grammar framework, showing a steady decline in accuracy and increased response times as proficiency levels advanced. These findings demonstrate that LLMs can successfully generate learning content, while highlighting the need for developers to strategically sequence question modalities to transition learners from passive recognition to active linguistic production.

**Keywords:** Grammar drill, EFL, log data, second language learning, CALL CEFR-J

## 1. Introduction

Second Language Acquisition (SLA) is being transformed by Computer-Assisted Language Learning (CALL) systems, which offer learners the benefit of unlimited, self-directed, and self-paced practice sessions anywhere at any time (Levy & Stockwell, 2013). However, developing these digital learning systems so they are effective and aligned to established frameworks can be difficult for educators and developers. Handcrafting learning content demands significant pedagogical expertise, making content creation both time-consuming and expensive. Consequently, traditional CALL software often relies on static, limited item banks; when students exhaust these resources, they face repetitive exercises that lead to boredom, diminished learning outcomes and motivation.

To overcome these scalability limitations, Large Language Models (LLMs) are increasingly used to automate content generation (Park & Derakhshan, 2026). LLMs can instantly generate vast, linguistically diverse, personalised, and contextually rich content. However, the rapid deployment of AI-generated content raises pedagogical concerns regarding proficiency alignment (Zhang & Huang, 2024). It remains unclear whether LLM-generated material reliably matches theoretical proficiency frameworks, or if it inadvertently introduces unintended friction for language learners. Furthermore, additional empirical research is required to evaluate how this generated content performs within localised English as a Foreign Language (EFL) contexts such as Japanese secondary education. Popular CALL apps (e.g., Duolingo) rely heavily on closed-ended question formats (e.g., multiple-choice (MCQ), drag-and-drop, and cloze tasks) to provide immediate feedback, yet few studies have compared the direct performative impacts of these distinct question modalities on actual students (Syahid, 2018).

To address these gaps, this study utilises *Grammar Gretel*, an EFL grammar drilling application driven by a dataset of over 55,000 LLM generated English-Japanese sentence pairs mapped to the CEFR-J framework. By analysing the log data of Japanese junior high school students, we evaluate the real-world viability of LLM-generated grammar drill practice. Specifically, this study investigates the following research questions:

1. How do different question modalities (MCQ, drag-and-drop, cloze) impact student accuracy and response times?
2. Do generated CEFR sentence levels accurately predict empirical task difficulty for learners?

## 2. Related Work

Within SLA research, the debate over the efficacy of explicit grammar drills and instruction versus implicit methods in EFL has been long-standing and complex (Şahinkaya, 2024). Historically, some researchers have argued that traditional, mechanical grammar drills are unnecessary and might even hinder natural language development, advocating instead for meaning-based and communicative approaches where learners are exposed to comprehensible input (Nassaji, 2016; Wong & Van Patten, 2003). Conversely, a robust body of comparative studies demonstrates that form-focused instruction and explicit grammar tasks frequently yield positive outcomes, enhancing both implicit and explicit grammatical knowledge and providing an essential structure for effective communication (Şahinkaya, 2024). More recently, researchers have highlighted the need to move beyond isolated declarative knowledge by applying concepts like Transfer-Appropriate Processing (TAP) (Nikouee & Oba, 2025). TAP suggests that to enable learners to use grammatical forms accurately and fluently in spontaneous communication, grammar practice should simulate the cognitive demands of real-world language use, thereby fostering procedural knowledge through meaningful, repeated practice (Nikouee & Oba, 2025).

When looking to automate this language practice, CALL applications frequently rely on fixed and structured exercise modalities, with the most common being MCQs, fill-in-the-blanks (cloze), pairing (vocabulary matching), and drag-and-drop exercises (Donati et al., 2024). Because traditional software has historically struggled to generate and evaluate natural language input beyond a predetermined list of messages or rules, developers lean heavily on these closed-ended formats to provide immediate and reliable feedback to learners. While it is well-documented that popular CALL applications rely heavily on these question types (Katinskaia, 2025; Ouyang et al., 2024; Shortt et al., 2021), there is a lack of empirical research comparing the direct cognitive and performative impacts of these different modalities on learners.

To streamline the creation of these platforms, app creators and researchers are increasingly generating grammar exercises with LLMs, moving away from a process that is traditionally highly labor-intensive and demands expert pedagogical knowledge to craft contextually relevant sentences and plausible distractors that successfully balance difficulty and maintain learner motivation (Donati et al., 2024; Vega et al., 2025). Early automated generation methods relied on rigid, rule-based algorithms to generate or extract sentences from text corpora, which often suffered from a lack of customisation for specific difficulty levels, topics, and creative variation (Donati et al., 2024). Recently, LLMs have emerged as powerful tools capable of autonomously generating accurate, diverse, contextually relevant, and self-contained sentences for grammar exercises, effectively overcoming the difficulties of previous rule-based systems (Katinskaia, 2025).

In the context of localised instruction, the CEFR-J framework in Japan provides a tailored version of the Common European Framework of Reference for Languages (CEFR); a globally recognised benchmark for assessing language competence through active "can-do" descriptors (Figueras, 2012). To better serve the distinct educational requirements, curricular guidelines, and learning obstacles of Japanese EFL students, this framework was adapted into the CEFR-J (Masashi, 2012). A core element of the CEFR-J framework is its highly granular classification of grammatical structures mapped directly to specific proficiency tiers. Evaluated from Japanese EFL textbook corpora, these grammatical items (e.g., present progressive; definite articles) are extracted via regular expressions (regex) to facilitate the automated identification of target grammatical features from raw text (Ishii & Tono, 2016). There remains a need for empirical validation of CEFR-J using learner data.

The current study utilises an LLM-generated corpus of more than 55,000 unique English-Japanese sentence pairs, tightly controlled for CEFR-J grammar points, to address two distinct gaps in the current literature. First, regarding question modality, while CALL applications rely heavily on closed-ended formats, few studies empirically compare the direct cognitive and performative impacts of these different mechanics on active learners (Ponce et al., 2021). Second, regarding CEFR-J validation, the framework categorises grammatical difficulty based primarily on textbook corpora; however, these theoretical tiers require empirical verification using real-world learner data (Tono, 2019).

## 3. Methodology

### 3.1 *Context and Participants*

This study was conducted at a high-performing junior high school in Japan. The participants consisted of 291 students in their second and third years of junior high school (age 14-15). Students interacted with *Grammar Gretel*, an EFL grammar drilling application, between January 13 and May 20 2026. The application was utilised both during formal class time and as an optional at-home study tool, and could be used on any device with a browser.

### 3.2 *System Design*

*Grammar Gretel* consists of three parts: a start page for users to configure their practice session, the interactive drill interface (Figure 1), and session end page showing the learner's mistakes, score, and scoreboard. On the start page, users can filter the question bank by school grade, CEFR difficulty level, topics, or specific parts of speech. To manage study pacing, the system allows students to cap their practice sets at 5, 10, 20, or 50 questions per session, while selecting the types of questions they want to practice. Text-To-Speech (TTS) functionality reads the sentence aloud automatically, and can be disabled.

#### 3.2.1 *Question Types and Distractors*

The system generates three question formats:

1. **MCQ:** The app blanks out a word. Students pick the correct answer from four choices (the target word and three distractors).
2. **Drag-and-Drop:** The system shows spaces to be filled with correct words and distractors provided. Students complete the sentence by dragging or tapping the correct words in order.
3. **Cloze:** Students type the missing grammar word into a blank. The app uses an auto-correct feature to accept minor spelling differences (like US vs. UK variations). In this mode, the Japanese translation is shown by default.

To choose which words to hide, the app first checks if a Part-of-Speech filter is active. If a student is practicing a specific word type (e.g. nouns or prepositions) the system cross-references the sentence with its pre-tagged grammar data and only allows those specific words to be hidden. If no filter is selected, or if the filter leaves the pool empty, the system defaults to making all words in the sentence available. Once the pool of allowed words is set, the system hides words based on the question type. In MCQ and cloze input modes, the system randomly selects one word from the allowed pool. It then separates any commas or periods from that word using regular expressions, attaching the punctuation to the surrounding text so that hints like a period do not give away the last word of the sentence. In drag-and-drop mode, the system sets an upper limit of hidden words, which is capped at half the sentence length. The app shuffles the allowed words and randomly selects a number of items up to that cap to serve as the blanks.

To create wrong answers (distractors) the system builds a word bank at the start of the session by gathering every word from all the sentences loaded for that drill and stripping their punctuation. For MCQs, the app selects three random words from this bank to sit alongside the correct answer. For drag-and-drop questions (Figure 1), the system selects four random words to act as distractors alongside the correct answers.

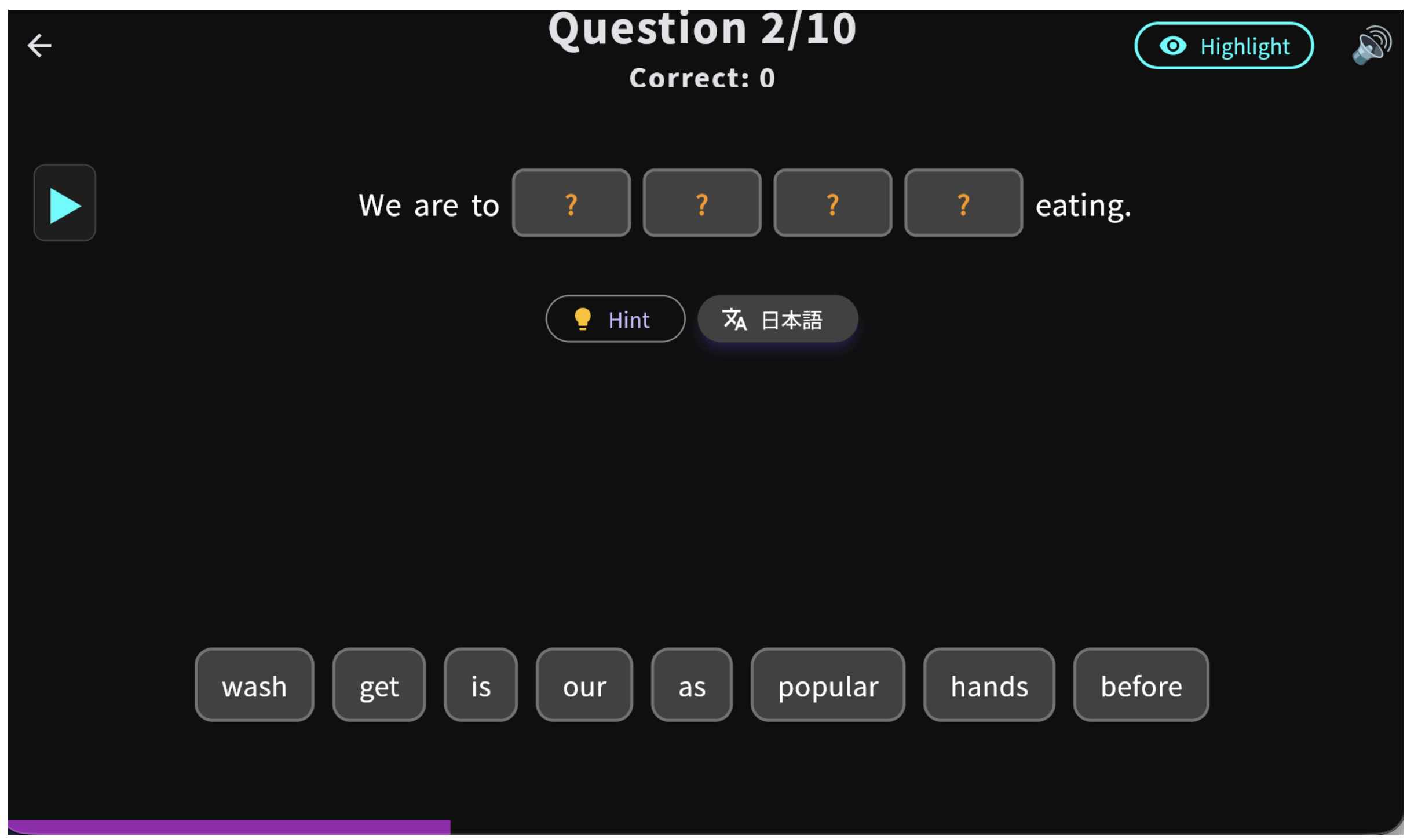

*Figure 1.* Drag and drop question example.

### 3.2.2 *Gamification*

To keep learners motivated, the app uses a countdown slider and a competitive scoring system. A countdown bar slides down at the bottom of the screen, automatically giving students more time for more time consuming question types (e.g., drag-and-drop), though students can also choose a slow, normal, or fast speed on the start page. Students earn points for accuracy and fast answers. However, each wrong guess and hint usage carries penalties. When the session ends, scores are shown on the leaderboard using fun, anonymous nicknames like "Jolly Puma" to encourage friendly competition without compromising privacy. To prevent mindless guessing, the app disallows question skipping until they have attempted to answer at least three times. Additionally, if the countdown timer runs out, the app pauses the screen, records a timeout, and displays the correct answer. The system then pauses to make sure the student is still paying attention, only advancing once the user manually clicks the *Next Question* button.

## 3.3 *Materials*

The sentences of *Grammar Gretel* are anchored to the CEFR-J framework, utilising its granular classification of grammatical structures to map items directly to specific proficiency tiers. A sample of these grammatical structures with examples is provided in Table 1. The complete repository of all 263 CEFR-J grammatical features (502 including negative forms) and their corresponding regex extraction formulas are available at cefr-j.org. CEFR-J describes grammar items between A1.1 and B2.2; C1 and C2 levels are not included. For grammar items with a CEFR range, the minimum level was utilised for analysis.

Table 1. *CEFR-J Grammar Item Examples*

| CEFR-J Level | Grammar Item | Example |
|---|---|---|
| A1.1 | INDEFINITE ARTICLES (a/an + noun) | I have a book. |
| A1.1-A1.2 | I am (... including question, negative) | I am happy. Am I next? I am not sad. |
| A1.1-B1.2 | WH- QUESTION: How ADJ/ADV ...? (How + adjective/adverb ...?) | How quickly can you run? |

| A1.2-A1.3 | TENSE/ASPECT: PRESENT PROGRESSIVE (present continuous) | She is reading. |
|---|---|---|
| A1.2-A2.2 | VERB to DO (verb + to-infinitive) | I want to sleep. |

To provide extensive, varied practice without repetition, a corpus of approximately 55,000 unique English-Japanese sentence pairs was generated using GPT-4.1-mini. For every grammar point in the CEFR-J taxonomy, the LLM generated batches of exactly 10 sentences across 11 distinct contextual domains (e.g., *Self, Family, & Friends*; *Health, Body, & Feelings*). This ensured a broad lexical variety while tightly controlling the target grammatical structure and intended CEFR difficulty level. At the start page, students choose between practising sentences created by their teacher, or free practice with this generated set. This study focuses exclusively on free practice with the LLM-generated sentences.

## 3.4 *Data Preprocessing*

To ensure analysis reflected genuine learning rather than system artifacts, responses recorded as system timeouts were removed. To eliminate extreme outliers from idle sessions, the upper 5% tail of attempts with long response times was trimmed. The app's gamified leaderboard and timer layout may have inadvertently incentivised rapid, mindless guessing for some learners. As shown in Figure 2, the response time distribution for MCQs is unique compared to the smoother, unimodal distributions of the drag-and-drop and cloze questions. The initial MCQ peak captures rapid-fire clicks occurring too quickly for learners to ready and process the question text.

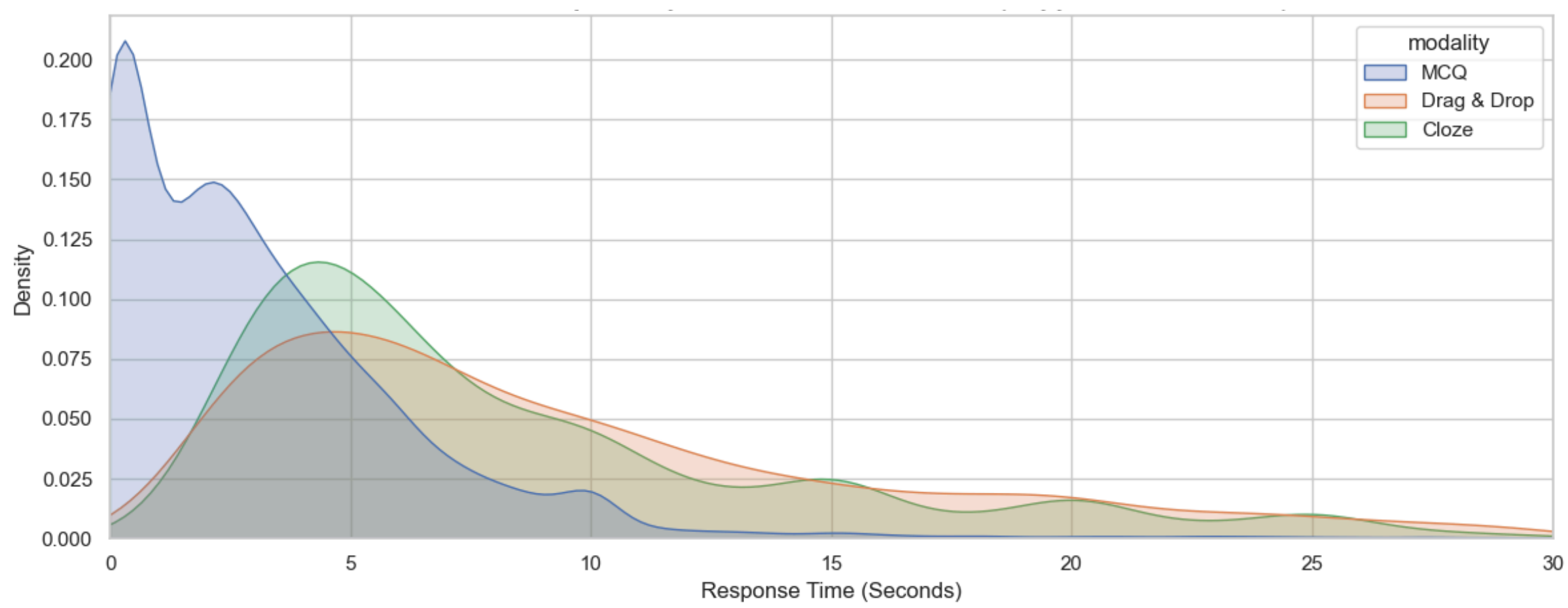

*Figure 2.* Density of Response Times (smoothed and capped at 30 seconds)

To filter out this rapid guessing behaviour, a Gaussian Kernel Density Estimate (KDE) was fitted to the MCQ response times across a high-resolution grid (0.1 to 10.0 seconds). By identifying the first local minimum of the density function, the optimal cutoff boundary was mathematically determined as $t$ = 1.04 seconds. All MCQ responses prior to 1.04 seconds were classified as rapid guessing and excluded from subsequent analyses: 5,662 MCQ in total. Only three responses in drag-and-drop and cloze questions were less than 1 second.

# 4. Results

The cleaned dataset includes log records from 291 unique students who answered a total of 29,963 questions. Across all practice sessions, learners achieved a global first-try accuracy rate of 71.4%, which rose to a global eventual accuracy rate of 90.4% after repeated attempts within the interface. The global median response time was 5.3 seconds. Table 2 provides a breakdown of total attempts across proficiency tiers and modalities.

Table 2. *Total Questions Attempted by CEFR Level and Modality*

| CEFR Level | MCQ | Drag & Drop | Cloze | Total |
|---|---|---|---|---|
| A1.1 | 4,614 | 2,859 | 3,255 | 10,728 |
| A1.2 | 3,059 | 1,865 | 2,150 | 7,074 |
| A1.3 | 1,575 | 982 | 1,088 | 3,645 |
| A2.1 | 1,548 | 1,121 | 1,381 | 4,050 |
| A2.2 | 500 | 264 | 336 | 1100 |
| B1.1 | 878 | 453 | 503 | 1834 |
| B1.2 | 252 | 156 | 220 | 628 |
| B2.1 | 326 | 152 | 170 | 648 |
| B2.2 | 140 | 48 | 68 | 256 |
| **Total** | **12,892** | **7,902** | **9,171** | **29,963** |

*RQ1: How do different question modalities (MCQ, drag-and-drop, cloze) impact student accuracy and response times?*

Analysing student accuracy and response latencies across the three interaction modalities revealed a clear performance hierarchy.

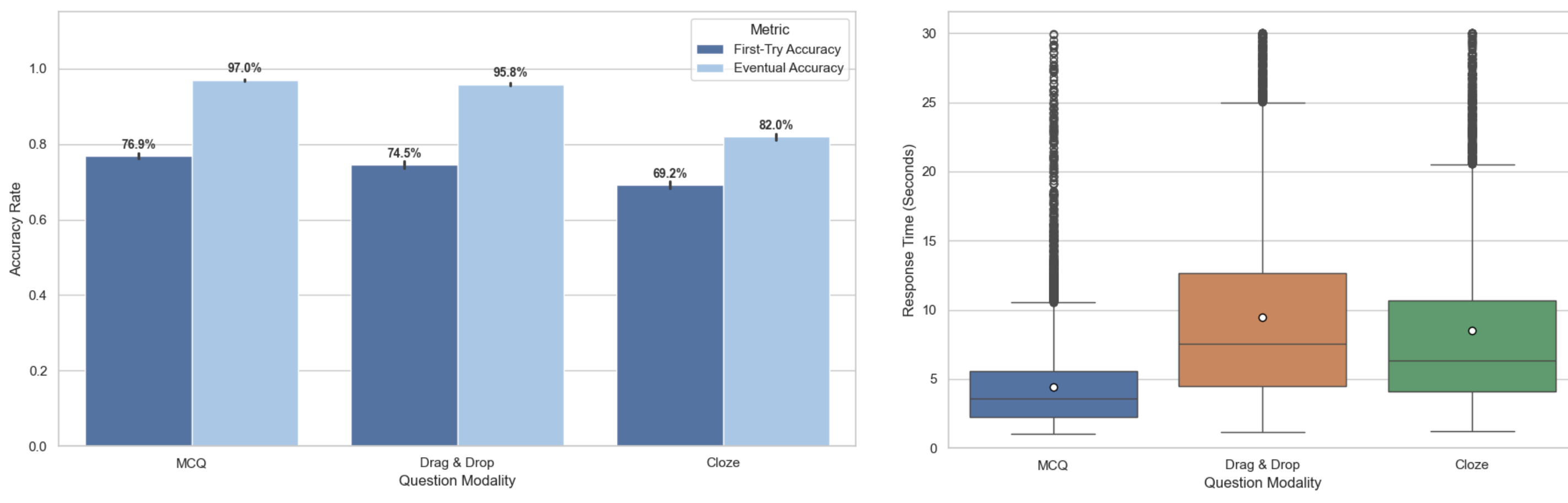

*Figure 3.* Accuracy by Question Modality

*Figure 4.* Response Time by Question Modality

As illustrated in Figure 3, MCQs achieved the highest baseline performance, with a first-try accuracy of 76.9% that rose to an eventual mastery rate of 97.0% through repeated attempts. drag-and-drop questions followed closely, yielding a 74.5% first-try accuracy rate and a 95.8% eventual accuracy rate. In contrast, cloze tasks presented the most substantial cognitive barrier, recording the lowest initial performance at 69.2% first-try accuracy and reaching an eventual accuracy floor of 82.0%. This sharper drop-off in eventual success indicates that the active recall required by production-based typing tasks is significantly harder for students to self-correct than recognition- or selection-based modalities.

The distribution of production latencies, detailed in Figure 4, reveals that task mechanics drastically alter response times. MCQs exhibited the lowest overhead, centering around a tight median response time of approximately 3.6 seconds, with a mean (indicated by the white dot) below 5 seconds. Conversely, drag-and-drop tasks incurred the heaviest temporal penalty; the median response time jumped past 7.5 seconds, with a mean approaching 10 seconds. This distribution reflects the dual burden of cognitive syntax reconstruction and the physical interaction time required to rearrange sentence chunks. cloze questions showed an intermediate latency profile, with a median response time of roughly 6.3 seconds and a mean near 8.5 seconds. Though faster to physically execute than drag-and-drop, cloze items maintained a broad distribution of elevated response times, reinforcing that retrieving specific lexical forms without contextual choices demands prolonged processing.

*RQ2: Do generated CEFR sentence levels accurately predict empirical task difficulty for learners?*

Mapping first-try accuracy and median response times across the proficiency spectrum shows that while the CEFR-J framework acts as an effective predictor on the whole, deviations emerge at specific thresholds.

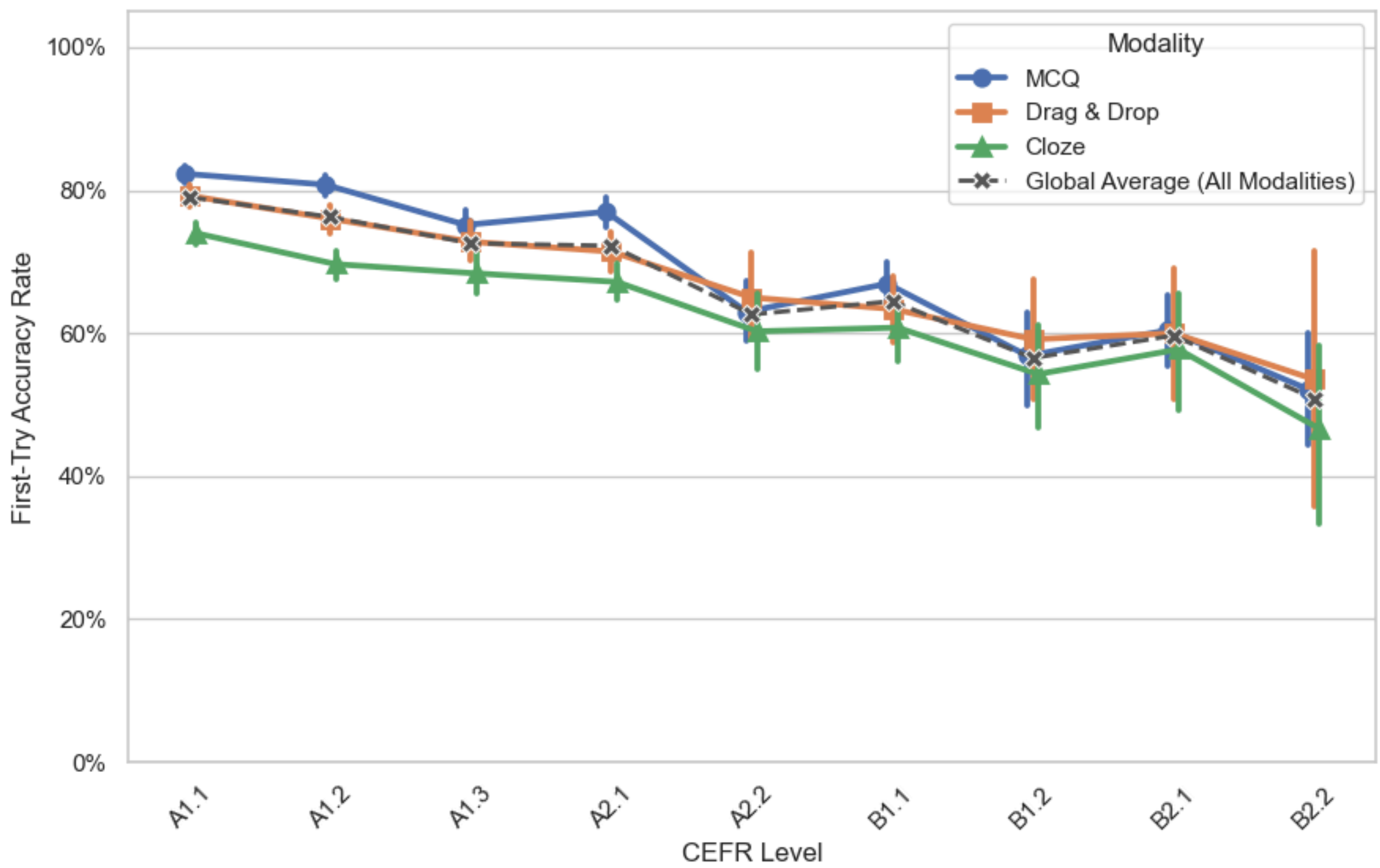

*Figure 5.* Accuracy across CEFR Levels

As shown in Figure 5, global average accuracy declines predictably across the beginner milestones, dropping from 79.13% at A1.1 to 72.67% at A1.3. Following a minor plateau at A2.1 (72.35%), accuracy falls to a floor of 50.88% at B2.2. Concurrently, as detailed in Figure 6, global median response times steadily rise from 4.32 seconds at A1.1 to a peak latency of 7.16 seconds at B1.2. Interestingly, overall response times actually decreased at the higher tiers, dropping to 6.26 seconds at B2.1 and further contraction to 6.04 seconds at B2.2. While the general pairing of decreasing success and prolonged processing across the early-to-mid tiers validates the core hierarchy of the CEFR tracking system, cross-modality analysis uncovers notable micro-level deviations. For instance, MCQ accuracy drops sharply at the A2.2 tier (63.06%), dipping below the theoretically more advanced B1.1 tier (67.02%). Furthermore, response latencies for drag-and-drop violate the linear trend entirely; median response times oscillate dramatically, experiencing severe spikes at A2.2 (10.69s) and B2.2 (14.83s), while contracting at B1.1 (10.03 s) and B2.1 (9.28s).

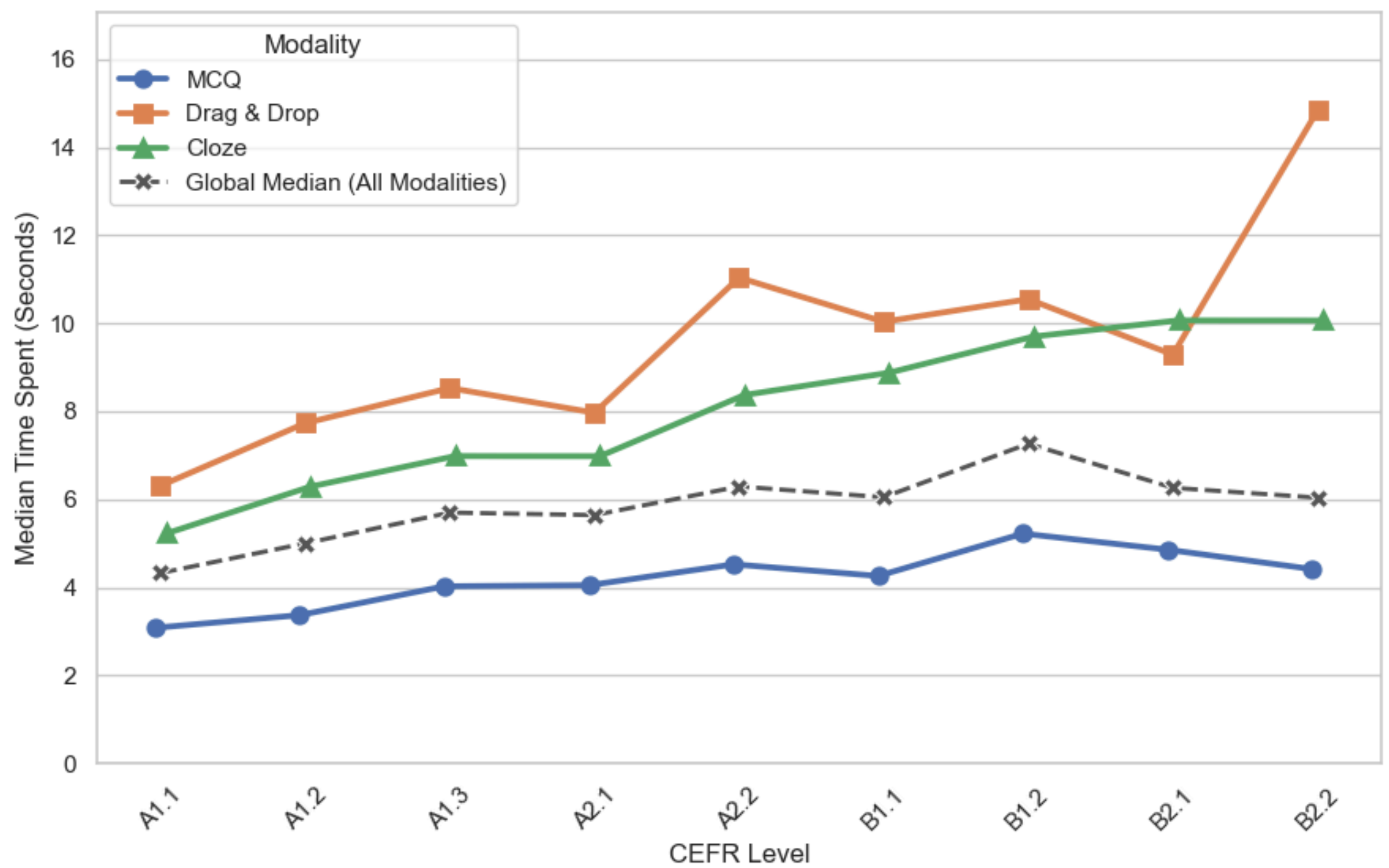

*Figure 6.* Response Time across CEFR Levels

## 5. Discussion

The objective of this study was to evaluate the pedagogical viability of LLM-generated grammar practice content within localised EFL contexts. By analysing interaction logs from Japanese junior high school students using *Grammar Gretel*, we aimed to determine how different question modalities impact learner performance (RQ1) and whether theoretical CEFR-J difficulty levels accurately predict empirical learner difficulty (RQ2).

### 5.1 *Modality Demands and Cognitive Processing (RQ1)*

Our findings reveal a clear performance and latency hierarchy among the three question types. MCQs achieved the highest first-try accuracy (76.9%) and the shortest median response time (3.6s). This aligns with recognised SLA literature indicating that recognition-based tasks carry lower cognitive loads because the correct answer is explicitly presented among distractors (Donati et al., 2024). Conversely, cloze questions presented the most formidable barrier to student success, yielding the lowest first-try accuracy (69.2%) and a relatively low eventual mastery ceiling (82%). This sharp drop-off in a learner's ability to self-correct highlights the unique difficulty of active, production-based recall. Without visual choices to prompt memory, typing a missing word forces the student to rely completely on internalised knowledge. Drag-and-drop questions occupied a unique middle ground: they achieved high accuracy (74.5%) but incurred the most severe temporal penalty, pushing median response times past 7.5 seconds. This extended latency reflects a dual burden: the physical interaction time required to drag interface blocks, and the deeper syntax reconstruction required to arrange sentence chunks into a coherent structure. When viewed through the lens of TAP, these results suggest that while MCQs are highly efficient for rapid vocabulary or form recognition, drag-and-drop and cloze questions are important for driving procedural language acquisition (Nikouee & Oba, 2025). Drag-and-drop questions force learners to think about word order and grammatical forms, whereas cloze tasks demand productive recall. Consequently, CALL app developers should not rely solely on MCQs for high gamified scores; rather, they should intentionally sequence these modalities to transition students from passive structural recognition to active linguistic production.

## 5.2 *Empirical Validation of the CEFR-J Framework (RQ2)*

Our data broadly validates the predictive power of the CEFR-J hierarchy. From the beginner milestones (A1.1) through intermediate thresholds (B2.2), average student accuracy declined predictably (from 79.13% to 50.88%). Global median response times similarly trended upward through the beginner and lower-intermediate levels, peaking at 7.16 seconds at B1.2, before decreasing at B2.1 (6.26s) and B2.2 (6.04s). This steady progression confirms that the textbook-derived, regex-parsed grammatical tiers of the CEFR-J index real-world difficulty for Japanese secondary school learners. It also provides evidence that LLMs can successfully generate sentence banks that preserve these targeted structural hierarchies. However, our analysis revealed notable anomalies that do not fit a perfectly linear difficulty curve, characterised by performance drops and severe latency fluctuations at specific thresholds. While overall response times decreased at B2.1 and B2.2, the drag-and-drop modality was an outlier, exhibiting response time spikes at A2.2 (10.69s) and B2.2 (14.83s). These anomalies could be due to structural inflection points in Japanese curriculum textbooks, or an artifact of using the minimum level where the CEFR-J framework provides a range. Crucially, the extreme latency spike observed at B2.2 is likely influenced by sample size limitations; B2.2 accumulated the lowest student engagement across the dataset, yielding just 48 total drag-and-drop attempts; amplifying data volatility. Furthermore, more complex grammatical structures generally result in longer sentences. When the LLM generated sentences at these advanced CEFR levels, the resulting items were naturally longer. In the drag-and-drop modality, a longer sentence drastically increases the number of potential word-chunk permutations. More words to drag simply take more time, perhaps explaining the 14.83 second drag-and-drop response time peak observed at B2.2, even as overall processing times for other modalities leveled off or decreased.

## 5.3 *Limitations and Future Work*

While this study leverages a large dataset of almost 30,000 question attempts, generalisability is limited by its sample population: junior high school students from a single, high-performing school in Japan. Additionally, the system's filtering feature introduces selection bias, as higher-proficiency or highly motivated students likely self-selected into advanced CEFR tiers while others remained at the A1 level. This confounding factor makes it difficult to decouple empirical task difficulty from student motivation profiles. This sample focussed on the lower levels of CEFR-J and more data is required at B1 and B2. Additionally, our distractors relied on purely random word extraction. Future investigations should refine automated content generation by implementing pedagogically driven distractor selection algorithms (such as BERT or parts-of-speech filtering to ensure distractors share the same semantic or syntactic category as the target word). Longitudinal analyses are also required to determine if sustained exposure to mixed-modality, LLM-generated practice yields lasting improvements in spontaneous communicative proficiency. Three key avenues for future research emerge:

- **Content Validation:** Although average accuracy validates the general CEFR-J taxonomy trajectory, future work could use automated CEFR levelers (e.g., (Uchida & Negishi, 2025) to audit the LLM-generated corpus and verify whether sentences match their intended difficulty. Future studies could also directly compare student performance on teacher-crafted versus LLM-generated sentences to evaluate the pedagogical efficacy and naturalness of automated content.
- **Practice Context and Dynamics:** Because Grammar Gretel was used both in class and at home, future analyses could isolate these usage logs to investigate how different environments alter processing latencies and accuracy. Additionally, modeling how first-try accuracy, eventual accuracy, and response times develop sequentially across individual sessions will reveal patterns regarding learner fatigue, cognitive friction, and short-term learning gains from immediate feedback.

- **Grammar Item-level Analysis and SLA Diagnostics:** A deeper item-level exploration is needed to identify which specific CEFR-J grammar items (e.g., definite articles vs. relative pronouns) EFL students find easiest to master versus those with which they most struggle. Mapping these empirical difficulties will allow us to cross-reference our data with historical SLA literature on common L1-Japanese specific errors. Ultimately, scaling this methodology to analyse error variations across diverse student backgrounds could allow gamified drilling applications to automate the holistic assessment of targeted grammatical difficulties for various native language (L1) groups acquiring English.

## Acknowledgements

This research is supported by the Council for Science, 3rd SIP JPJ012347, JSPS Grant-in-Aid for Scientific Research (B) JP23H01001, (A) JP23H00505, and KAKENHI Grant Number 26KJ1488.

## References

Donati, N., Periani, M., Natale, P. D., Savino, G., & Torroni, P. (2024). *Generation and evaluation of English grammar multiple-choice cloze exercises*. 325–334.

Figueras, N. (2012). The impact of the CEFR. *ELT Journal*, *66*(4), 477–485.

Ishii, Y., & Tono, Y. (2016). A frequency survey of grammar items for the CEFR-J grammar profile. *Proceedings of the 22nd Meeting of the Association for Natural Language Processing*, 777–780.

Katinskaia, A. (2025). An overview of artificial intelligence in computer-assisted language learning. In *arXiv [cs.CL]*. arXiv. https://doi.org/10.48550/arXiv.2505.02032

Levy, M., & Stockwell, G. (2013). *CALL dimensions: Options and issues in computer-assisted language learning*. Routledge.

Masashi, N. (2012). The development of the CEFR-J: Where we are, where we are going. *New Perspectives for Foreign Language Teaching in Higher Education: Exploring the Possibilities of Application of CEFR. Tokyo: Tokyo University of Foreign Studies*, 10–116.

Nassaji, H. (2016). Research Timeline: Form-focused instruction and second language acquisition. *Language Teaching*, *49*(1), 35–62.

Nikouee, M., & Oba, T. (2025). Beyond drills: Optimizing grammar practice for transfer. *TESL Canada Journal*. https://teslcanadajournal.ca/index.php/tesl/article/view/1841

Ouyang, Z., Jiang, Y., & Liu, H. (2024). The effects of Duolingo, an AI-integrated technology, on EFL learners' willingness to communicate and engagement in online classes. *The International Review of Research in Open and Distributed Learning*, *25*(3), 97–115.

Park, Y., & Derakhshan, A. (2026). From Gutenberg to the classroom: large-scale generation and validation of vocabulary-controlled EFL reading materials. *Language Testing in Asia*, *16*(1).

Ponce, H. R., Mayer, R. E., & Loyola, M. S. (2021). Effects on test performance and efficiency of technology-enhanced items: An analysis of drag-and-drop response interactions. *Journal of Educational Computing Research*, *59*(4), 713–739.

Şahinkaya, H. A. (2024). Explicit vs. implicit grammar teaching in EFL classroom: A literature review. *International Journal of Academic Research in Education*, *9*(1), 14–26.

Shortt, M., Tilak, S., Kuznetcova, I., Martens, B., & Akinkuolie, B. (2021). Gamification in mobile-assisted language learning: a systematic review of Duolingo literature from public release of 2012 to early 2020. *Computer Assisted Language Learning*, 1–38.

Syahid, A. (2018). Usability of Moodle question types by EFL teachers. *Proceedings of International Conference on English Language Teaching (INACELT)*, *2*(1), 224–237.

Tono, Y. (2019). Coming Full Circle —From CEFR to CEFR-J and back. *CEFR Journal - Research and Practice 1*. https://doi.org/10.37546/jaltsig.cefr1-1

Uchida, S., & Negishi, M. (2025). Assigning CEFR-J levels to English learners' writing: An approach using lexical metrics and generative AI. *Research Methods in Applied Linguistics*, *4*(2), 100199.

Vega, J., Rodriguez, M., Check, E., Moran, H., & Loo, L. (2025). Duolingo evolution: From automation to artificial intelligence. In *Communications in Computer and Information Science* (pp. 54–71). Springer Nature Switzerland.

Wong, W., & Van Patten, B. (2003). The Evidence is IN: Drills are OUT. *Foreign Language Annals*, *36*(3), 403–423.

Zhang, Z., & Huang, X. (2024). The impact of chatbots based on large language models on second language vocabulary acquisition. *Heliyon*, *10*(3), e25370.